\documentclass[%
 reprint,
superscriptaddress,
 amsmath,amssymb,
 aps,
]{revtex4-1}

\usepackage{graphicx}
\usepackage{hyperref}
\usepackage[version=4]{mhchem}
\usepackage{float}
\usepackage{caption}
\usepackage{subcaption}
\usepackage{tikz}

\hypersetup{
	colorlinks   = true,
	citecolor    = blue,
	linkcolor = blue,
    urlcolor = blue
}

\usepackage{color}

\begin{document}
\title{Magnetism and Piezoelectricity in Stable Transition Metal Silicate Monolayers}
\author{Kayahan Saritas}
\affiliation{Department of Applied Physics, Yale University, New Haven, CT, 06520}
\author{Nassar Doudin}
\affiliation{Department of Chemistry, Yale University, New Haven, CT, 06520}
\author{Eric I. Altman}
\affiliation{Department of Chemistry, Yale University, New Haven, CT, 06520}
\author{Sohrab Ismail-Beigi}
\affiliation{Department of Applied Physics, Yale University, New Haven, CT, 06520}
\email{sohrab.ismail-beigi@yale.edu}

\begin{abstract}
Two-dimensional van der Waals (2D vdW) single layered materials with ferromagnetism and piezoelectricity have been the subject of recent attention. Despite numerous reports of 2D ferromagnetic materials, developing an air stable and transferable vdW material has been challenging. To address this problem, we studied layered transition metal silicates that are derivatives of kaolinites and lizardites with transition metals substituting on Al$^{3+}$ and Mg$^{2+}$ sites using {\it ab initio} Density Functional Theory (DFT) calculations. This class of materials are appealing because they meet the symmetry requirements for piezoelectricity and can host a range of transition metal cations. As oxides, these materials are inherently stable in air. Following our previous experimental work, we predict that these compounds are stable under varying O$_2$ partial pressure and can be synthesized using a surface assisted method. We also show that the oxidation states of the substituted transition metal ions can be tuned through the level of hydrogenation.
\end{abstract}

\maketitle

\section{Introduction}
Two-dimensional van der Waals (2D VDW) materials that display ferromagnetism, piezoelectricity, and ferroelectricity have received increased attention \cite{Burch2018a, Gibertini2019, Gong2017, C9CP00972H}. 
VDW stacking of multiple 2D layers with these complementary properties can help develop multifunctional materials \cite{Cheng2016, Novoselovaac9439, Geim2013}. 
Despite the fact that there are various well-studied piezoelectric 2D materials available \cite{Blonsky2020, Duerloo2012}, developing an air-stable and transferable VDW single layered material whose ferromagnetism can be tuned under electric or elastic field has been challenging.
Air stability is an important problem in this regard as it presents significant challenges in isolating and studying the single layers \cite{Huang2017a}. 
Single layers of CrI$_3$ \cite{Huang2017a}, Cr$_2$Ge$_2$Te$_6$ \cite{Tian_2016} and FePS$_3$ \cite{Wang_2016} are also shown to be ferromagnetic, but similarly they suffer from sensitivity to oxidation. In comparison, the search space for ferroic oxide VDW layers remains under-explored. Oxides have the advantage of being stable under ambient conditions, e.g., most metals spontaneously form a thin layer of metal oxide on their surfaces \cite{Cabrera_1949}. 

\begin{figure*}
\centering
    \begin{tikzpicture}
    \node[inner sep=0pt] {\includegraphics[width=\linewidth]{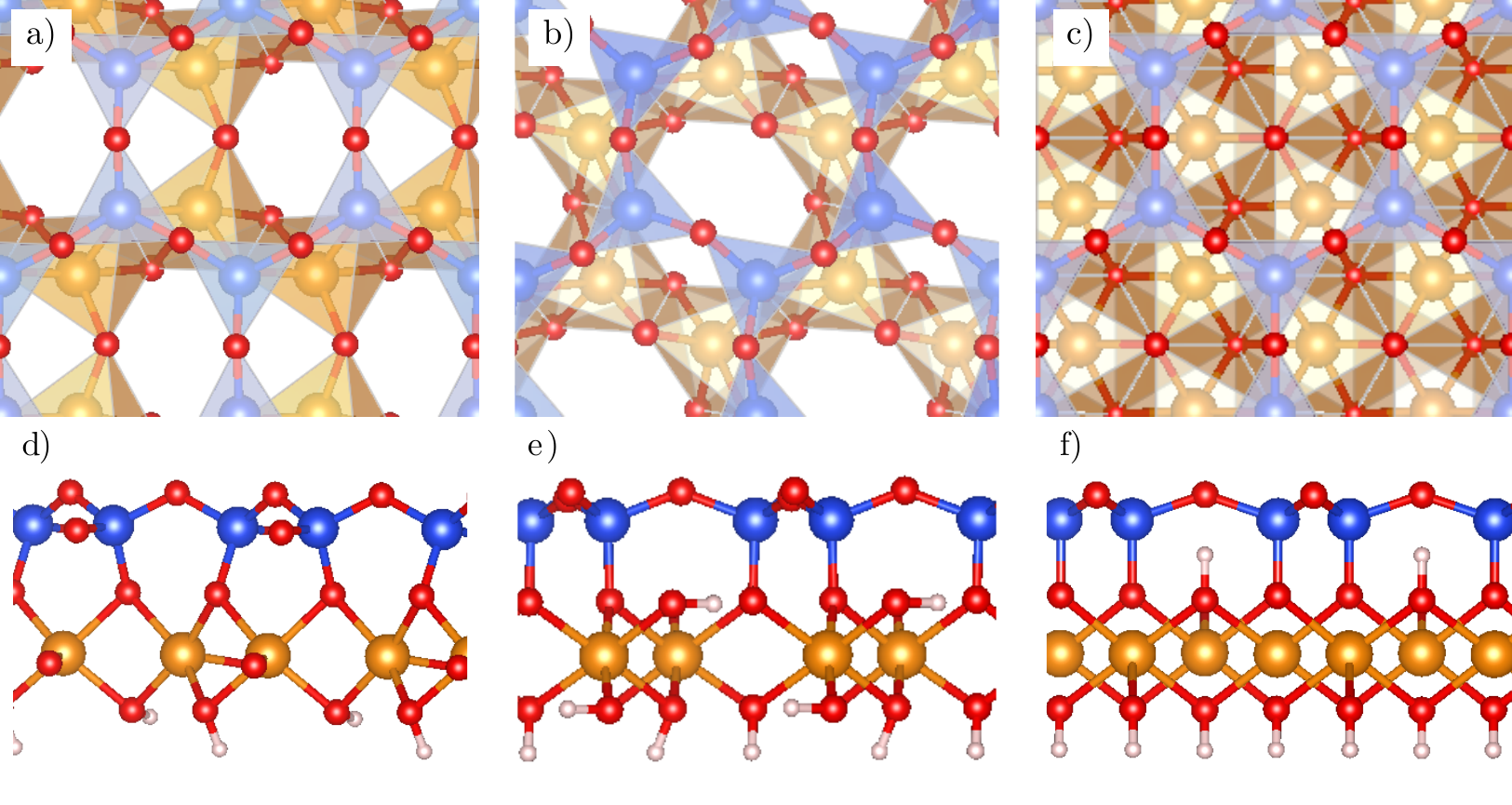}};
        \draw[->, line width=0.5mm] (-8.5,0) -- (-7.5,0);      
        \draw[->, line width=0.5mm] (-8.5,0) -- (-8.5,1);      
        \node[align=center] at (-8.5, 1.2) {\bf \large y};
        \node[align=center] at (-7.3, 0) {\bf \large x};

        \draw[->, line width=0.5mm] (-2.5,0) -- (-1.5,0);      
        \draw[->, line width=0.5mm] (-2.5,0) -- (-2.5,1);      
        \node[align=center] at (-2.5, 1.2) {\bf \large y};
        \node[align=center] at (-1.3, 0) {\bf \large x};

        \draw[->, line width=0.5mm] (3.8,0) -- (4.8,0);      
        \draw[->, line width=0.5mm] (3.8,0) -- (3.8,1);      
        \node[align=center] at (3.8, 1.2) {\bf \large y};
        \node[align=center] at (5, 0) {\bf \large x};

        \draw[->, line width=0.5mm] (-8.5,0) -- (-7.5,0);      
        \draw[->, line width=0.5mm] (-8.5,0) -- (-8.5,1);      
        \node[align=center] at (-8.5, 1.2) {\bf \large y};
        \node[align=center] at (-7.3, 0) {\bf \large x};

        \draw[->, line width=0.5mm] (-8.5,-4) -- (-8.5,-3);      
        \node[align=center] at (-8.5, -2.7) {\bf \large z};
        
        \draw[->, line width=0.5mm] (-2.5,-4) -- (-2.5,-3);      
        \node[align=center] at (-2.5, -2.7) {\bf \large z};
        
        \draw[->, line width=0.5mm] (3.8,-4) -- (3.8,-3);      
        \node[align=center] at (3.8, -2.7) {\bf \large z};
    \end{tikzpicture}
  \caption{Top view of the structural templates of the isolated 2D silicate layers studied in this work: a) nontronite-like, b) kaolinite-like, and c) lizardite-like transition metal silicates. Respective side views are indicated in d), e) and f). Blue, red and orange indicate Si, O and transition metal atom sites, respectively. Blue polyhedra highlight Si-O polyhedra, and gold polyhedra surround the transition metals. In side images, hydrogen atoms are indicated in white for the highest hydrogenation possible in their respective structures.}
  \label{fgr:top_view}
\end{figure*}
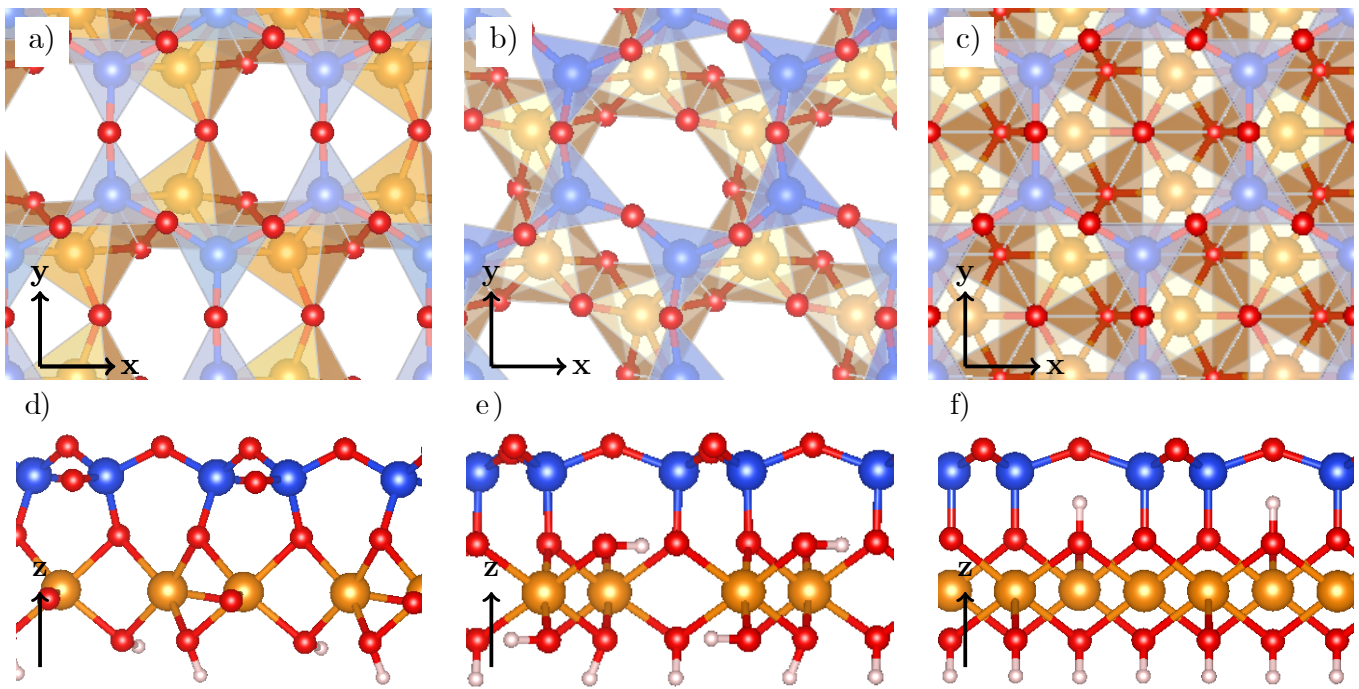

Transition metal silicate sheets, which have been grown on metal substrates via annealing at elevated temperatures \cite{Wodarczyk2013, Zhou2019, Fischer2015}, are air stable. Thanks to the open-shell transition metal atoms, their magnetic properties can also be tailored. Growing a 2D transition metal silicate starts with depositing Si or SiO and the transition metal ion, such as Ti \cite{Fischer2015}, Fe \cite{Wodarczyk2013} and Ni \cite{Zhou2019},  at modest temperatures, which is then followed by annealing above 950 K. In all these cases the transition metal silicate thermodynamically competes with the formation of bilayer SiO$_2$ \cite{Altman2013, C7CP02382K, Li2017, Lichtenstein2012, Shaikhutdinov2013}. However, parameters such as annealing time, temperature, oxygen pressure, Si and transition metal coverage can facilitate metastable phase formation.
As 2D Ti-silicate \cite{Fischer2015} can be grown on metal substrates, using these experimental methods it should be possible to synthesize first row transition metal silicates films with smaller transition metal ions (considering the relatively large  size of the Ti atom) with a judicious choice of substrate. 
The resulting transition metal silicates resemble crystal structures of naturally existing sheet silicates (phyllosilicates), particularly that of dehydroxylated nontronite, M$_2$Si$_2$O$_8$ \cite{Wodarczyk2013, Tissot2016} as in Fig. \ref{fgr:top_view}a. The competing bilayer SiO$_2$ would be composed of a six-membered ring of SiO$_4$ tetrahedra with out-of-plane mirror symmetry. In the nontronite case, however, the transition metal polyhedra still form six-membered rings, but they are rotated in such a way so that all the polyhedra are five-fold coordinated. In nontronite, therefore, the rings of metal polyhedra have four edge-sharing and two corner-sharing connections.

Even though nontronite-like 2D transition metal silicates are synthesized on metal substrates, it is possible that closely related crystals, such as kaolinite \cite{Bish1993} and lizardite \cite{Mellini1987} which are also phyllosilicates, can also coexist under  similar thermodynamic conditions. 
Kaolinite and Lizardite have the chemical formulae  Al$_2$Si$_2$O$_9$H$_4$ and Mg$_3$Si$_2$O$_9$H$_4$, hence their dehydroxylated forms are Al$_2$Si$_2$O$_9$ and Mg$_3$Si$_2$O$_9$, as shown in Fig. \ref{fgr:top_view}b-c and e-f.
It is known that transition metal atoms such as Fe, Ni, and Co can almost fully substitute Al and Mg in these crystals and similar phyllosilicates \cite{Coey1982, Bayliss1981, Manceau1998, dainyak2006, MALDEN1967, WOODWARD2018270}. 
Greenalite ((Fe$^{2+}$,Fe$^{3+}$)$_{2-3}$Si$_2$O$_9$H$_4$) \cite{Shirozu1965} and nepouite/pecoraite  (Ni$_3$Si$_2$O$_9$H$_4$) \cite{Brindley1975, C2TA00257D} crystals correspond to the Fe- and Ni-substituted kaolinite and lizardite. 
Magnetic properties of greenalite were  previously studied and \textit{intrasheet} ferromagnetic order was observed \cite{Ballet1978}. It was argued that 90$^{\circ}$ Fe$^{2+}$-O-Fe$^{2+}$ interactions lead to net magnetisation in the 2D layer \cite{Ballet1978}. 
Therefore, there is a large chemical space to be explored that could be engineered for 2D ferromagnetism. Additionally, the crystalline space groups of kaolinite and lizardite break inversion symmetry, thus these compounds are automatically piezo-active \cite{Parkhomenko1971}. 

In this work, we used density functional theory with Hubbard-$U$ correction (DFT+$U$) to study structural, energetic, magnetic, electronic and piezoelectric properties of 2D transition metal silicates.  We have studied the derivatives of nontronite M$_2$Si$_2$O$_8$H$_x$, kaolinite M$_2$Si$_2$O$_9$H$_x$ and lizardite M$_3$Si$_2$O$_9$H$_x$ as isolated 2D layers in vacuum at various degrees of hydrogenation, where M=Cr, Mn, Fe, Co, Ni and $x$=0-4 ($x$=0-2 for nontronites). Each transition metal derivative is investigated systematically. For each compound, we report oxidation states of the transition metals, formation and hull energies, energy differences between antiferromagnetic and ferromagnetic phases and average magnetic moments. Additional data describe the stability regions for compounds that are on the convex hull as well as Gibbs free energies of hydrogenation. Finally, piezoelectric properties are reported for the thermodynamically stable and some metastable layers. 
We find that there is a rich chemical space for transition metal silicates that are thermodynamically stable with finite piezoelectricity and potential for a magnetic phase with net magnetization. 

\section{Methods}
All DFT calculations were performed using the Vienna ab initio Simulation Package (VASP) \cite{Kresse1996, Kresse1996a}, using the Perdew-Burke-Ernzerhof (PBE) \cite{Perdew1996} exchange-correlation functional along with the Hubbard-$U$ approach (PBE+$U$) \cite{Dudarev1998}. We employ $U$ values of 3.7 eV for Cr, 3.9 eV for Mn, 5.3 eV for Fe, 3.32 eV for Co and 6.2 eV for Ni for all transition metal oxides using the guidelines in Materials Genome Project \cite{Jain2011, Jain2013}. 
These $U$ values were determined according to their accuracy in reproducing the formation energies of all known binary metal oxides \cite{Jain2011}.
Elemental compounds, whose energies are needed for constructing phase diagrams, are calculated using PBE. 
PBE and PBE+$U$ energies are combined using an established mixing scheme calibrated using binary oxide formation energies \cite{Jain2011}: the PBE and PBE+$U$ mixing scheme provides compositional correction parameters for each atom. Similar composition-based corrections are widely used to correct DFT energies to construct phase diagrams \cite{Saal2013, Stevanovic2012}.

To construct the phase diagrams, we use a procedure based on prior work by \citet{Persson2012}. To determine the chemical potentials of any compound $i$ under standard conditions, $\mu_i^0$, we define reference chemical potentials, $\mu_{ref}^0$, such that $\mu_i^0 = g_i^0 - \mu_i^{ref}$, where $g_i^0$ is the Gibbs free energy of the species $i$ under standard conditions. Gibbs free energy is defined as $g_i^0=h_i^0-Ts_i^0$, where $h$ and $s$ are enthalpy and entropy respectively. Thus, if the reference is at standard conditions, we can simplify as $\mu_i^0 = h_i^0 - h_i^{ref}$ and here, $h_i^{ref} = E_i^{0,DFT}$, where $E_i^{0,DFT}$ the DFT ground-state energy for the elemental solid. Phase diagrams in this work are constructed using elemental solids, solid oxides, oxygen, hydrogen and water, which are then used to calculate the relative stabilities of transition metal silicates. Due to the difficulties in treating such a  broad range of materials accurately with a single choice of DFT functional, we resort to several practical approximations and empirical corrections to obtain $\mu_i^{ref}$ for each compound. 
For elemental solids, we approximate as $\mu_i^{ref} \approx E_i^{0, DFT}$, hence $s_i(T)=0$. For oxygen gas, however, the reference chemical potential is defined as $\mu_O^{ref} = E_O^{0, DFT} + \Delta{E^{corr}_O} - Ts^{ref}_O$ where  $\Delta{E^{corr}_O}$ is the oxygen gas correction term added to DFT calculations to better reproduce experimental binary oxide formation energies \cite{Wang2007}. The reference, $ref$, here depends on the pressure/temperature of the gases.
We use the entropy $s^{ref}_O$ from prior work \citet{kubaschewski1993materials}. For a binary solid oxide, $A_xO_y$, like the elemental solids, we approximate the entropic terms as negligible \cite{Persson2012} and write the chemical potential for that compound as $\mu_{A_xO_y}^{0} = E_{A_xO_y}^{DFT} - {x}\mu^{ref}_A - {y}\mu^{ref}_O$. Water is a particularly difficult case for most theoretical methods. Therefore, we use the free energy of formation of \ce{H2O} at room temperature, $\mu_{H_2O}^{0}$ = -2.46 eV/H$_2$O. The hydrogen chemical potential depends on the chemical potential of H$_2$O and O, hence it is obtained indirectly as follows. Using the water formation reaction, we define $\mu_H^{ref} = 1/2[g^{ref}_{H_2O}-\mu^{ref}_O-\mu_{H_2O}^{0}]$. Here, $g^{ref}_{H_2O} = E^{0, DFT}_{H_2O}-Ts_{H_2O}^0$. Numerical values for the variables discussed here are provided in the Supplementary Information along with a more detailed explanation and examples.

We calculate elastic tensor coefficients, $C_{ij}$, with no ionic relaxations, using the finite differences method and the strain coefficients of the piezoelectric tensor, $e_{ij}$, via the Berry phase method \cite{King-Smith1993, Resta2007} in VASP. An orthogonal supercell was used to calculate the elastic and piezoelectric constants. A gamma-centered reciprocal space grid of 4x8x1 (corresponding to a grid density of 120 \AA$^{-3}$) for sampling the first Brillouin zone and an energy cutoff of 520 eV are used in all our calculations. We ensured a spacing of minimum 30 {\AA} of vacuum between the periodic images of layers in all calculations. Dipole corrections \cite{Neugebauer1992} were also included along the out-of-plane direction to reduce spurious interactions between periodic images. 

\section{Results and discussion}
Using the phase diagram generation procedure described in the Methods section, we study the stabilities of all the transition metal silicates (from Cr to Ni) shown in Fig. \ref{fgr:top_view} at varying hydrogenation levels. We consider three structural templates for these transition metal silicates. These templates are based on natural minerals or are ones that have been synthesized.  These are nontronite-like silicates M$_2$Si$_2$O$_8$H$_{n}$ where $n=0-2$, kaolinite-like silicates M$_2$Si$_2$O$_9$H$_{n}$ where $n=0-4$, and lizardite-like silicates M$_3$Si$_2$O$_9$H$_{n}$ where $n=0-4$. For any hydrogenation level, $n$, all possible combinations of hydrogen bindings to the binding sites were considered, and the minimum energy structure is reported. In Fig. \ref{fgr:top_view}d)-f), these structures are shown with the highest degree of hydroxylation possible. Kaolinite has two additional oxygens compared to nontronite, and all transition metal polyhedra in kaolinite are six-fold coordinated and edge-sharing. Lizardite, on the other hand, has one additional metal atom compared to kaolinite, which makes the metal oxide layer triangular, as opposed to the honeycomb lattice in kaolinite. Once the structural stabilities are explained (section~\ref{sec:thermostability}, we describe other computed physical properties (remaining sections).

\subsection{Thermodynamic Stabilities}
\label{sec:thermostability}
\subsubsection{Fe-silicates}
In Table \ref{table:fe-stability}, we show the average oxidation number, transition metal electronic configuration, formation and hull energies and average Fe magnetic moments for all the Fe-silicates studied. Average oxidation numbers, $N_{ox}$, are determined by  assuming O and Si to be closed shell ions (i.e., O$^{2-}$ and $Si^{4+}$ ions).  The $N_{ox}$ can be used with the transition metal electronic configuration (EC) to show that a charge ordered structure is found. EC was determined by using the magnetic moment on the transition metal atoms and also chemical intuition. For example, for Fe$_3$Si$_2$O$_9$ with $N_{ox}=3.33+$ and EC of $d^4,d^5$ indicates that one of the Fe atoms is $d^4$ (4+) and the other two are $d^5$ (3+). We find that this structure is monoclinic ($Cm$, $\#$8) with $\gamma$=119.64$^\circ$. However, when the same structure is forced to have trigonal symmetry ($P31m$, $\#$157), the energy is increased by 0.05 eV/f.u., and all three Fe atoms become magnetically identical (as expected when constrained to be magnetically collinear). Similar symmetry breaking and charge disproportionation is observed in all structures with non-integer $N_{ox}$ in Table \ref{table:fe-stability}. Structural parameters of all the 2D materials are given in SI \cite{Supp}. 

\begin{table}[ht]
\caption{Average oxidation number of the Fe atoms ($N_{ox}^{Fe}$), Fe electronic configuration (EC), calculated formation enthalpies (E$_f$ in eV/atom) and energy above the convex hull (E$_{hull}$ in eV/atom) for the 2D iron silicates studed in this work. Average magnetic moment per Fe atom ($\mu_M$ in $\mu_B/$atom) in the high-spin FM state is provided. Compositions on the hull of the phase diagram are in bold font. 
\label{table:fe-stability}}

\begin{tabular}{l c c c c c }
\hline
\hline
Material & $N_{ox}^{Fe}$ & EC & E$_f$ & E$_{hull}$ & $\mu_M$\\
\hline
\multicolumn{6}{l}{\textit{Nontronites}} \\
Fe$_2$Si$_2$O$_8$	    & 4+ & d$^4$          & 	-2.253	&	0.175	&	2.19	\\
Fe$_2$Si$_2$O$_8$H	    & 3.5+ & d$^4$, d$^5$ & 	-2.236	&	0.122	&	2.56	\\
Fe$_2$Si$_2$O$_8$H$_2$	& 3+ & d$^5$          & 	-2.220	&	0.078	&	2.96	\\
\hline
\multicolumn{6}{l}{\textit{Kaolinites}}	\\													
Fe$_2$Si$_2$O$_9$	    & 5+         & d$^3$        &	-2.023	&	0.218	&		3.53	\\
Fe$_2$Si$_2$O$_9$H	    & 4.5+       & d$^3$, d$^4$ &	-2.027	&	0.163	&	 	3.82	\\
Fe$_2$Si$_2$O$_9$H$_2$	& 4+         & d$^4$        &	-2.039  &	0.106	&   	3.87	\\
Fe$_2$Si$_2$O$_9$H$_3$	& 3.5+       & d$^4$, d$^5$ &	-2.037	&	0.054	&		4.08	\\
{\bf Fe$_2$Si$_2$O$_9$H$_4$}	& 3+         & d$^5$        &	-2.042  &	0.000	&		4.35	\\
\hline
\multicolumn{6}{l}{\textit{Lizardites}}	\\													
Fe$_3$Si$_2$O$_9$	    & 3.33+ & d$^4$, d$^5$ &	-2.327	&	0.091	&	4.10	\\
Fe$_3$Si$_2$O$_9$H	    & 3+    & d$^5$        &	-2.292	&	0.066	&	4.30	\\
Fe$_3$Si$_2$O$_9$H$_2$	& 2.67+ & d$^5$, d$^6$ &	-2.184	&	0.050	&	4.03 \\
Fe$_3$Si$_2$O$_9$H$_3$	& 2.34+ & d$^5$, d$^6$ &	-2.095	&	0.023	&	3.93 \\
{\bf Fe$_3$Si$_2$O$_9$H$_4$}	& 2+    & d$^6$        &	-2.015	&	0.000	&	3.75	\\
\hline
\end{tabular}
\end{table}

The convex hull energies, $E_{hull}$, in Table \ref{table:fe-stability} are determined using a Fe-Si-O-H quaternary phase diagram \cite{Supp}. The only compounds that have zero hull energy in Table \ref{table:fe-stability} are  Fe$_2$Si$_2$O$_9$H$_4$ and Fe$_3$Si$_2$O$_9$H$_4$. Although Fe-nontronites were shown to exist \cite{Wodarczyk2013} on Ru (111) substrate, Table \ref{table:fe-stability} shows that this phase is unstable in isolation even when it is fully hydrogenated. Nevertheless, in all the compounds, there is a clear trend of decreasing hull energy with hydroxylation. However, it is possible that metastable compounds (e.g., $E_{hull}<50$ meV/atom) can be kinetically trapped making them experimentally accessible. In the literaure, a tolerance on the DFT hull energies of around 10 meV/atom is used to eliminate false negatives on the convex hull \cite{Narayan2016a}. However, the smallest non-zero hull energy in Table \ref{table:fe-stability} is 23 meV/atom, which is well above this likely tolerance of DFT error.

\begin{figure*}
\centering
\includegraphics[]{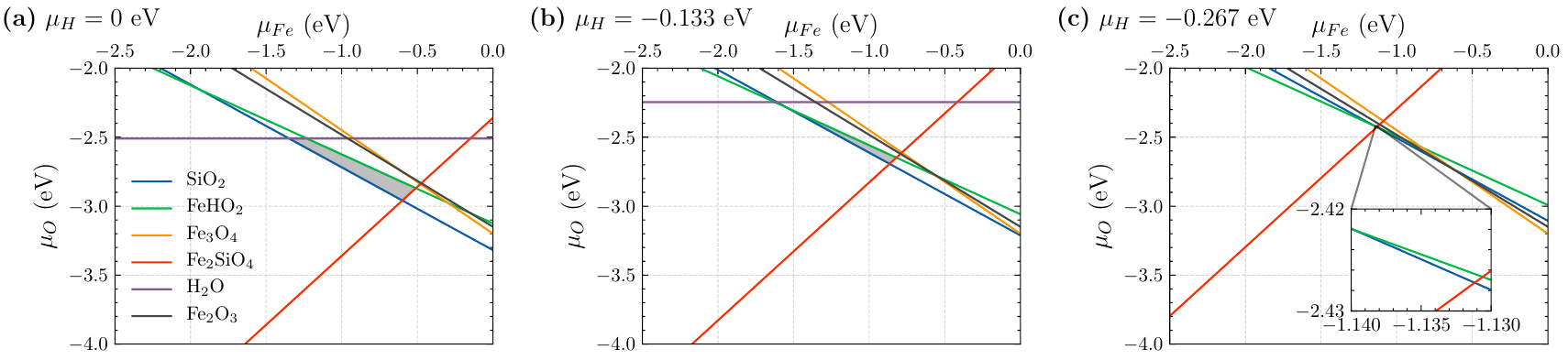}
\hskip -0ex
  \caption{Phase diagrams for 2D Fe$_3$Si$_2$O$_9$H$_4$ with a) hydrogen chemical potential $\mu_H$ is set to 0 eV, b) $\mu_H=-0.133$ eV and c) $\mu_H=-0.267$ eV. Stability regions of Fe$_3$Si$_2$O$_9$H$_4$ are shaded in gray. In c) the shaded area is given in the inset for further detail.}
  \label{fgr:fe3}
\end{figure*}

\begin{figure*}
\centering
\includegraphics[]{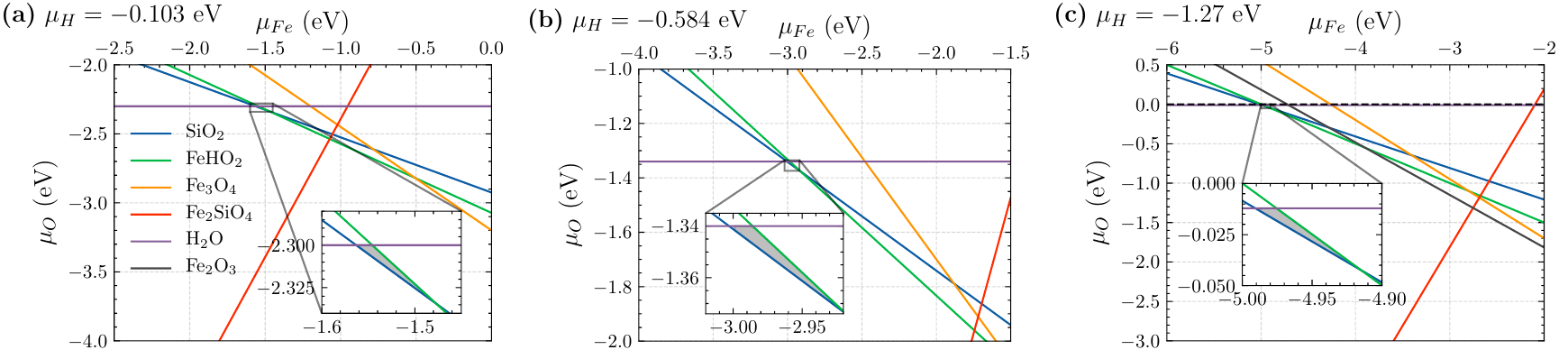}
  \caption{Phase diagrams for Fe$_2$Si$_2$O$_9$H$_4$: a) hydrogen chemical potential $\mu_H$ is set to -0.103 eV, b) $\mu_H$ is set to -0.584 eV and c) $\mu_H$ is set to -1.27 eV. Stability regions are shaded in gray. Insets show the shaded regions in greater detail.  }
  \label{fgr:fe2}
\end{figure*}

In Fig. \ref{fgr:fe3}, we show the chemical stability ranges for  Fe$_3$Si$_2$O$_9$H$_4$ as a function of the O and Fe chemical potentials, $\mu_{O}$ and $\mu_{Fe}$. 
We assume that our quaternary compound (here Fe$_3$Si$_2$O$_9$H$_4$) is in equilibrium with the reservoirs, so that there are three independent chemical potentials. We choose $\mu_{Fe}$ and $\mu_{O}$ as the two of the independent parameters for our study and fix the third,  $\mu_{H}$, at various values to understand how the stability regions are modified.  Three different $\mu_H$ values are selected based on the maximum, average and minimum $\mu_H$ where Fe$_3$Si$_2$O$_9$H$_4$ can exist based on the calculated phase diagram. In general, a "facet" of phase diagram using convex hull formalism indicates a hull member, or in other words it means an $N$-1 dimensional geometric domain on the convex hull where a set of compounds that can coexist in equilibrium. Here, $N$ is the number of compounds used to construct the phase diagram. In a binary convex hull phase diagram, for example, there are two compounds, hence the dimension of a facet would be one. In this case the facet has a special name called tie-line and each tie-line connects two points on the phase diagram which coexist on this tie-line. For a quaternary compound, such as Fe$_3$Si$_2$O$_9$H$_4$, a facet is then a three dimensional object which is formed by connecting four points in the convex hull phase diagram. These four points correspond to four compounds that coexist within the three dimensional convex hull facet. All the facets of the quaternary phase diagram where Fe$_3$Si$_2$O$_9$H$_4$ is found to be thermodynamically stable are given in the Supplementary Information (SI) \cite{Supp}. For example, $\mu_H$=0 eV is possible on a facet, where Fe$_3$Si$_2$O$_9$H$_4$-H$_2$-SiO$_2$-H$_2$O can coexist for example. However, $\mu_H$=-0.267 eV is obtained from the Fe$_3$Si$_2$O$_9$H$_4$-Fe$_2$SiO$_4$-FeHO$_2$-SiO$_2$ facet (coexistence region). Therefore, the stability of Fe$_3$Si$_2$O$_9$H$_4$ is bounded by $0 > \mu_H > -0.267$ eV.
With $\mu_H$ is fixed in Fig. \ref{fgr:fe3} and \ref{fgr:fe2}, the H$_2$O coexistence curves in Fig. \ref{fgr:fe3} are always horizontal. 
In Fig. \ref{fgr:fe2}, we perform the same analysis on Fe$_3$Si$_2$O$_9$H$_4$ and find that the stability region, where no other phase precipitates, is much smaller in all cases compared to Fig. \ref{fgr:fe3}.

There are various trends that can guide  experimental synthesis. Decreasing the hydrogen chemical potential yields a smaller stability region, hence hydrogen rich environments should more easily yield Fe$_3$Si$_2$O$_9$H$_4$. Next, we find that the stability region of Fe$_3$Si$_2$O$_9$H$_4$ is mainly controlled by competition of SiO$_2$ and FeHO$_2$ formation. In addition, phase separation into binary oxides is not relevant. Next, under fixed $\mu_{O}$, increasing the $\mu_{Fe}$, will lead to precipitation of FeHO$_2$, and decreasing $\mu_{Fe}$ (increasing $\mu_{Si}$) will lead to precipitation of SiO$_2$. Under fixed $\mu_{Fe}$, increasing $\mu_{O}$ (decreasing $\mu_{Si}$) will generally lead to formation of FeHO$_2$. In Fig. \ref{fgr:fe3}a, there is a small region near $\mu_{Fe}<-1.3$ eV, where increased $\mu_{O}$ will first lead to precipitation of H$_2$O and then FeHO$_2$ will follow. 

We compare the stability curves of Fe$_2$Si$_2$O$_9$H$_4$ in Fig. \ref{fgr:fe2} to the stability curves of Fe$_3$Si$_2$O$_9$H$_4$ in Fig. \ref{fgr:fe3}. Our main finding is that the stability region is much smaller compared to Fig. \ref{fgr:fe3}. In Fig. \ref{fgr:fe2}a, for example, it is a very small region around $\mu_{Fe}$=-1.5 eV and $\mu_{O}$=-2.3 eV.
We believe the rather small sizes of the stable regions in both figures is primarily due to the fact that both the Fe$_3$Si$_2$O$_9$H$_4$ and Fe$_2$Si$_2$O$_9$H$_4$ are isolated 2D layers, but their energetics are competing against bulk compounds.  The iron silicate can be further stabilized trivially by considering its bulk version, but more relevantly by placing it on a substrate for added stabilization from the wetting (binding) energy to the substrate.
 
For example in Fig. \ref{fgr:fe3}, such an added binding energy would increase the stability region as the SiO$_2$ and Fe$_2$SiO$_4$ curves would shift down with the increased formation energy of the Fe$_3$Si$_2$O$_9$H$_4$ while the curves for all the other iron oxides and FeHO$_2$ remains fixed at constant $\mu_H$. The same effect must also occur for Fe$_2$Si$_2$O$_9$H$_4$ given the fact that SiO$_2$ curve will shift down as a result. 

\begin{figure}[ht]
    \centering
    \includegraphics[width=\linewidth]{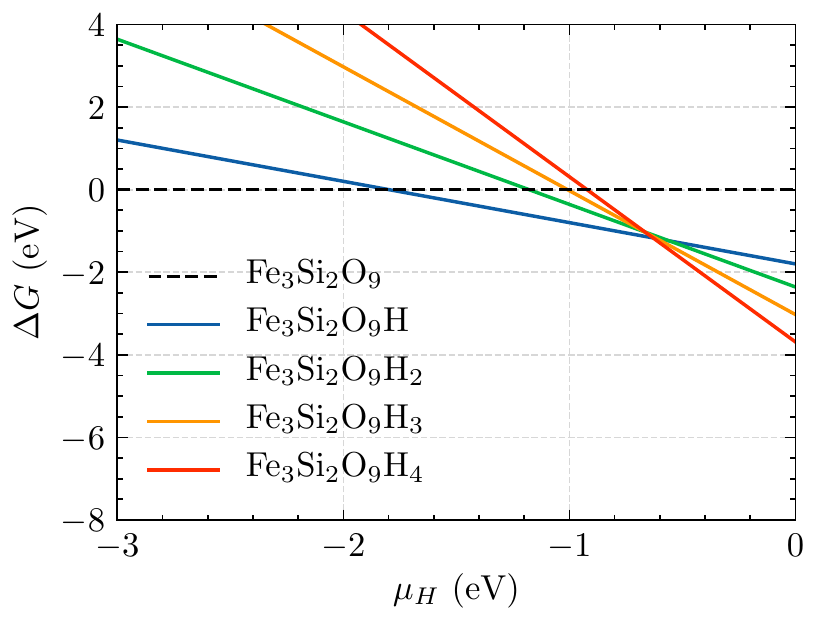}
    \caption{Gibbs free energy, $\Delta{G}$, of the hydrogenation reaction of Eq.~(\ref{eq:gibbs}) for Fe$_3$Si$_2$O$_9$H$_n$ as a function of the chemical potential of hydrogen, $\mu_H$.}
    \label{fgr:fe3h}
\end{figure}

\begin{figure}[ht]
\centering
    \includegraphics[width=\linewidth]{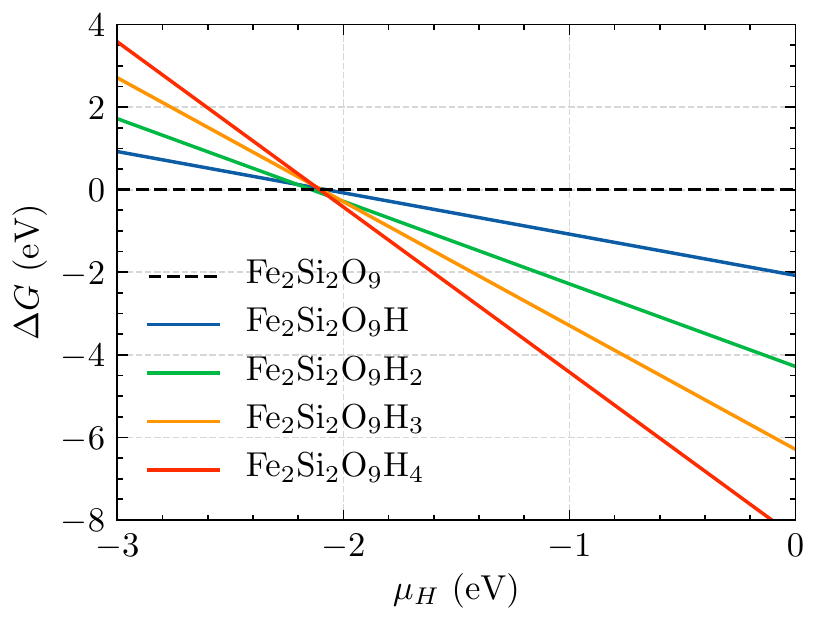}
  \caption{Gibbs free energy, $\Delta{G}$, for Eq.~(\ref{eq:gibbs}) and Fe$_2$Si$_2$O$_9$H$_n$ as a function of the chemical potential of hydrogen.}
  \label{fgr:fe2h}
\end{figure}

In Fig. \ref{fgr:fe3h} and  and \ref{fgr:fe2h}, we show the Gibbs free energy of hydrogenation for the following chemical reaction,
\begin{equation}
    \ce{M_xSi2O9} + \frac{n}{2}\ce{H2} \longrightarrow \ce{M_xSi2O9H_n}
    \label{eq:gibbs}\,.
\end{equation}
We note that these curves only indicate the relative stability of each compound, not the thermodynamic stability of the compound which is determined through hull energies in Table \ref{table:fe-stability}. Equation \ref{eq:gibbs} only describes whether hydrogenation of the transition metal silicates is thermodynamically favorable.

Among the hydroxylated phases of Fe$_3$Si$_2$O$_9$H$_n$, only \ce{Fe3Si2O9H} and \ce{Fe3Si2O9H4} are relatively stable compared to the others which can be explained through the valence state of the Fe atoms. In \ce{Fe3Si2O9H}, all the Fe atoms are in the 3+ state, whereas in \ce{Fe3Si2O9H4} all the Fe atoms are in 2+ state. Hence, we can conclude that under increasingly H-rich conditions, the charge disproportionation of the Fe atoms is expected to disappear. 

In Fig. \ref{fgr:fe2h}, we show the same analysis for Fe$_2$Si$_2$O$_9$H$_n$ and find that, essentially, a transition only occurs between Fe$_2$Si$_2$O$_9$ and Fe$_2$Si$_2$O$_9$H$_4$ compounds. This can be understood in a similar way as in \ref{fgr:fe3h}. In Fe$_2$Si$_2$O$_9$, all Fe atoms are in the 5+ state, and in Fe$_2$Si$_2$O$_9$H$_4$ they reduce to the 3+ state. Given that Fe would prefer oxidation states with 2+ and 3+, there is no intermediate hydrogenated compound that is stable. Additionally, in Fig. \ref{fgr:fe2h}, the transition occurs at a similar chemical potential to the first transition in Fig. \ref{fgr:fe3h}, but much lower than the second transition in Fig. \ref{fgr:fe3h} where complete hydrogenation has occurred. This implies that it is significantly easier to obtain a completely hydrogenated 2D Fe-silicate with the kaolinite-like structure once the base compound is formed. 

Separately, we calculate the cleavage energy of 2D Fe-silicates to show that these 2D materials are exfoliable assuming that their bulk counterparts can be synthesized. For Fe-kaolinite we find the cleavage energy to be 14 meV/{\AA}$^2$, while for Fe-lizardites the cleavage energy is 18 meV/{\AA}$^2$. According to the large-scale screening studies of the exfoliation energies of 2D compounds \cite{Mounet2018}, exfoliation energies below 30-35 meV/{\AA}$^2$ are classified as ``easily exfoliable'' regime, meaning that it can be exfoliated using simple techniques such as mechanical exfoliation. For the rest of the transition metal silicates, we provide these numbers in the SI \cite{Supp}, but overall they are between 18-22 meV/{\AA}$^2$ for lizardites and 14-18 meV/{\AA}$^2$ for kaolinites, showing that the transition metal only has negligible effect on the cleavage energies.

\subsubsection{Cr-silicates}
Our analysis of Cr-silicates and the remaining the silicates will follow closely our analysis of Fe-silicates. 
In Table \ref{table:cr-stability}, we present our data on Cr-silicates in the same manner as the Fe-silicates in Table \ref{table:fe-stability}. Hull energies in Table \ref{table:cr-stability} are determined using a Cr-Si-O-H quaternary phase diagram using the compounds listed in the supplementary information \cite{Supp}. The only stable compound in Table \ref{table:cr-stability} is Cr$_2$Si$_2$O$_9$H$_4$. There is a clear trend of increasing stability with increased hydrogenation in nontronites and kaolinites, but for lizardites the trend is in the opposite direction. This may be because, in lizardites, increased hydrogenation yields the 2+ charge state Cr, which is energetically unfavorable compared to the  3+ state. This agree with the fact that Cr$^{3+}$ compounds are more common than Cr$^{2+}$ compounds \cite{chem_elements}.

\begin{table}[ht]
\caption{Ground state properties of 2D Cr-silicates. \label{table:cr-stability}}
\begin{tabular}{l c c c c c}
\hline
\hline
Material & $N_{ox}^{Cr}$ & EC & E$_f$ & E$_{hull}$ & $\mu_M$ \\
\hline
\multicolumn{6}{l}{\textit{Nontronites}} \\
Cr$_2$Si$_2$O$_8$	    & 4+ & d$^2$          & -2.553	&	0.057	&	2.19 \\
Cr$_2$Si$_2$O$_8$H	    & 3.5+ & d$^2$, d$^3$ &	-2.428	&	0.113	&	2.55 \\			
Cr$_2$Si$_2$O$_8$H$_2$	& 3+ & d$^3$          &	-2.312	&	0.17	&	2.96 \\			
\hline
\multicolumn{6}{l}{\textit{Kaolinites}}	\\													
Cr$_2$Si$_2$O$_9$	    & 5+   & d$^1$        & -2.368	&	0.041	&	1.43	\\		
Cr$_2$Si$_2$O$_9$H	    & 4.5+ & d$^1$, d$^2$ & -2.315	&	0.045	&	1.94	\\		
Cr$_2$Si$_2$O$_9$H$_2$	& 4+   & d$^2$        &	-2.280	&	0.036	&	2.32	\\	
Cr$_2$Si$_2$O$_9$H$_3$	& 4.5+ & d$^2$, d$^3$ &	-2.235	&	0.019	&	2.71    \\
{\bf Cr$_2$Si$_2$O$_9$H$_4$}	& 3+   & d$^3$        &	-2.201	&	0	    &	2.96	\\	
\hline
\multicolumn{6}{l}{\textit{Lizardites}}	\\													
Cr$_3$Si$_2$O$_9$	    & 3.33+      & d$^2$, d$^3$ & 	-2.621	&	0.032	&	2.83	\\
Cr$_3$Si$_2$O$_9$H	    & 3+         & d$^3$        &	-2.545	&	0.046	&	2.98	\\		
Cr$_3$Si$_2$O$_9$H$_2$	& 2.67+      & d$^3$, d$^4$ &	-2.346	&	0.083	&	3.36 \\	
Cr$_3$Si$_2$O$_9$H$_3$	& 2.34+      & d$^3$, d$^4$ &	-2.170	&	0.116	&	3.68 \\
Cr$_3$Si$_2$O$_9$H$_4$	& 2+         & d$^4$        &	-2.027	&	0.132	&	3.75 \\	
\hline
\hline
\end{tabular}
\end{table}

The stability diagram for Cr$_2$Si$_2$O$_9$H$_4$ is provided in the supplementary information. Similar to Fe$_2$Si$_2$O$_9$H$_4$ in Fig. \ref{fgr:fe2}, the stability region of Cr$_2$Si$_2$O$_9$H$_4$ is bounded by H$_2$O, SiO$_2$ and CrHO$_2$. When constructing Cr-Si-O-H phase diagram, our search did not yield a thermodynamically stable Cr-Si-O ternary compound, unlike Fe$_2$SiO$_4$. The corresponding Cr$_2$SiO$_4$ structure has been synthesized only at elevated temperatures with rapid quenching \cite{Dollase1994}, and it is reported to be metastable within DFT \cite{osti_1193501}, suggesting that the structure can only be kinetically trapped. We find that the Gibbs free energy of hydrogenation of Cr$_2$Si$_2$O$_9$H$_4$, given in supplementary information, is similar to the hydrogenation of Fe$_2$Si$_2$O$_9$H$_4$ in Fig. \ref{fgr:fe2h}. The main difference between the hydrogenation of two materials is that the energetic crossing between Cr$_2$Si$_2$O$_9$ and Cr$_2$Si$_2$O$_9$H$_4$ occurs at a larger chemical potential compared to Fe$_2$Si$_2$O$_9$H$_4$. This agrees with the fact that Fe has a larger ionization potential compared to Cr, hence hydrogenation is comparatively more favorable at lower hydrogen availability.

\subsubsection{Mn-silicates}
Table \ref{table:mn-stability} shows that none of the Mn-silicates are found to be stable in the Mn-Si-O-H phase diagram we constructed using the compounds given in supplementary information. Nevertheless, similar trends are observed as in Cr and Fe-silicates, such that increased hydrogenation typically stabilizes the compound.  This trend is more obvious in Mn$_2$Si$_2$O$_9$H$_n$, where Mn transitions from being 5+ to 3+ with increased hydrogenation. However, for Mn$_3$Si$_2$O$_9$H$_n$, cases where Mn has an integer oxidation state of 3+ or 2+ are more stable compared to fractional oxidation states. We find that Mn$_3$Si$_2$O$_9$H$_n$ favors the 3+ oxidation state only slightly more than the 2+ state, which is expected given that these two oxidation states are the most commonly observed for Mn compounds \cite{chem_elements}.

\begin{table}[ht]
\caption{ Ground state properties of 2D Mn-silicates \label{table:mn-stability}}

\begin{tabular}{l c c c c c }
\hline
\hline
Material & $N_{ox}^{Mn}$ & EC & E$_f$ & E$_{hull}$ & $\mu_M$\\
\hline
\multicolumn{6}{l}{\textit{Nontronites}} \\
Mn$_2$Si$_2$O$_8$ 	    & 4+   & d$^3$          & 	-2.375	&	0.172	& 3.28	\\
Mn$_2$Si$_2$O$_8$H	    & 3.5+ & d$^2$, d$^3$   &	-2.317	&	0.125	& 3.51	\\
Mn$_2$Si$_2$O$_8$H$_2$	& 3+   & d$^4$          &	-2.256	&	0.095	& 3.85	\\
\hline
\multicolumn{6}{l}{\textit{Kaolinites}}	\\													
Mn$_2$Si$_2$O$_9$	    & 5+   & d$^2$        &	-2.122	&	0.169	&	3.02	\\
Mn$_2$Si$_2$O$_9$H	    & 4.5+ & d$^2$, d$^3$ &	-2.158	&	0.114	&	3.06	\\
Mn$_2$Si$_2$O$_9$H$_2$	& 4+   & d$^3$        &	-2.172	&	0.032	&	3.16	\\
Mn$_2$Si$_2$O$_9$H$_3$	& 3.5+ & d$^3$, d$^4$ &	-2.126	&	0.018	&	3.54	\\
Mn$_2$Si$_2$O$_9$H$_4$	& 3+   & d$^4$        & -2.069	&	0.013	&	3.87	\\
\hline
\multicolumn{6}{l}{\textit{Lizardites}}	\\													
Mn$_3$Si$_2$O$_9$	    & 3.33+         & d$^3$, d$^4$  & 	-2.467	&	0.038	&	3.92	\\
Mn$_3$Si$_2$O$_9$H	    & 3+            & d$^4$         &   -2.391	&	0.026	&	3.93	\\
Mn$_3$Si$_2$O$_9$H$_2$	& 2.67+         & d$^4$, d$^5$  & 	-2.274	&	0.046	&	4.27	\\
Mn$_3$Si$_2$O$_9$H$_3$	& 2.34+         & d$^4$, d$^5$  & 	-2.168	&	0.045	&	4.44	\\
Mn$_3$Si$_2$O$_9$H$_4$	& 2+            & d$^5$         & 	-2.088	&	0.031	&	4.63	\\
\hline
\end{tabular}
\end{table}

We studied the hydrogenation of Mn-silicates using Eq. \ref{eq:gibbs} and show the associated figures in the supplementary information. We find the same relation between the Gibbs free energy of hydrogenation in Mn$_3$Si$_2$O$_9$H$_n$ and Fe$_3$Si$_2$O$_9$H$_n$ in Fig. \ref{fgr:fe3h}. The only difference is that the hydrogen chemical potentials at the  transition points for Mn$_3$Si$_2$O$_9$H$_n$ are higher compared to Fe$_3$Si$_2$O$_9$H$_n$, which is a similar trend we noted for Cr-silicates.  For Mn$_2$Si$_2$O$_9$H$_n$, however, there is an additional regions of hydrogen chemical potential where Mn$_3$Si$_2$O$_9$H and Mn$_3$Si$_2$O$_9$H$_2$ are stable as well. This is in contrast to the Fe-silicates where we have a direct transition from Fe$_3$Si$_2$O$_9$ to Fe$_3$Si$_2$O$_9$H$_4$ in Fig. \ref{fgr:fe2h}. This may be because Mn tends to commonly accept a wider range of oxidation states compared to Fe, therefore the differences between higher degrees of ionization energies should be small enough to allow stepwise hydrogenation as opposed to the Fe-silicate examples.

\subsubsection{Co-silicates}
Table \ref{table:co-stability} shows that the stability trends for Co-silicates are very similar to those in Table \ref{table:fe-stability}. Again, the stability increases with increased hydrogenation and hull energies of trioctahedral Co-silicates are consistently smaller than dioctahedral derivatives. Similarly, we find that the stability regions of Co$_3$Si$_2$O$_9$H$_4$ are larger than Co$_2$Si$_2$O$_9$H$_4$ (see  the supplementary information). An important difference between the Co and Fe-silicates is that Co$_2$Si$_2$O$_9$H$_4$ starts forming at a lower hydrogen chemical potential than for Fe$_{(2,3)}$Si$_2$O$_9$H$_4$ silicates. Given that the stability region is mainly determined by SiO$_2$, CoHO$_2$ and H$_2$O curves, this can be explained by the fact that  the CoHO$_2$  formation enthalpy is significantly higher than FeHO$_2$ (1.13 eV vs 1.56 eV). Compared to FeHO$_2$, CoHO$_2$ is less likely to form.

\begin{table}[ht]
\caption{ Ground state properties of 2D Co-silicates \label{table:co-stability}}

\begin{tabular}{l c c c c c}
\hline
\hline
Material & $N_{ox}^{Co}$ & EC & E$_f$ & E$_{hull}$ & $\mu_M$\\
\hline
\multicolumn{6}{l}{\textit{Nontronites}} \\
Co$_2$Si$_2$O$_8$	    & 4+    & d$^5$         & 	-2.078	&	0.091	& 2.52 \\
Co$_2$Si$_2$O$_8$H	    & 3.5+  & d$^6$,d$^5$   & 	-2.059	&	0.049	& 2.68 \\
Co$_2$Si$_2$O$_8$H$_2$  & 3+    & d$^6$	        & 	-2.028	&	0.028	& 3.11\\
\hline
\multicolumn{6}{l}{\textit{Kaolinites}}	\\													
Co$_2$Si$_2$O$_9$	    & 5+    & d$^4$         & 	-1.924	&	0.078	& 2.58\\
Co$_2$Si$_2$O$_9$H	    & 4.5+  & d$^4$, d$^5$  & 	-1.865	&	0.092	& 2.25\\
Co$_2$Si$_2$O$_9$H$_2$	& 4+    & d$^5$         & 	-1.868	&	0.050	& 2.05\\
Co$_2$Si$_2$O$_9$H$_3$	& 3.5+  & d$^5$, d$^6$  & 	-1.861	&	0.020	& 2.78\\
\bf{Co$_2$Si$_2$O$_9$H$_4$}	& 3+    & d$^6$         & 	-1.847	&	0.000	& 3.15\\
\hline
\multicolumn{6}{l}{\textit{Lizardites}}	\\													
Co$_3$Si$_2$O$_9$	    & 3.33+ & d$^5$         & 	-2.049	&	0.036	& 2.07\\
Co$_3$Si$_2$O$_9$H	    & 3+    & d$^5$, d$^6$  & 	-2.003	&	0.036	& 3.01\\
Co$_3$Si$_2$O$_9$H$_2$	& 2.67+ & d$^6$         & 	-1.958	&	0.033	& 2.99\\
Co$_3$Si$_2$O$_9$H$_3$	& 2.34+ & d$^6$, d$^7$  & 	-1.932	&	0.010	& 2.87\\
\bf{Co$_3$Si$_2$O$_9$H$_4$}	& 2+    & d$^7$         & 	-1.899	&	0.000	& 2.74\\
\hline
\end{tabular}
\end{table}

We studied the hydrogenation of Co-silicates using Eq. \ref{eq:gibbs} and show the  plots in the supplementary information. In Co$_3$Si$_2$O$_9$H$_4$, we find that the Gibbs free energy of hydrogenation has the following trend going from Fe to Mn and Co. In Fe$_3$Si$_2$O$_9$H$_n$ (Fig. \ref{fgr:fe3h}), the hydrogenation is stepwise such that after Fe$_3$Si$_2$O$_9$, first Fe$_3$Si$_2$O$_9$H and then Fe$_3$Si$_2$O$_9$H$_4$ is formed.  Mn$_3$Si$_2$O$_9$H$_4$ is similar to this but the range of stability for Mn$_3$Si$_2$O$_9$H is smaller compared to Fe$_3$Si$_2$O$_9$H. Following this trend, the range of stability for Co$_3$Si$_2$O$_9$H disappears completely (we will see that the range of stability for Ni$_3$Si$_2$O$_9$H is again smaller compared to Fe$_3$Si$_2$O$_9$H and Mn$_3$Si$_2$O$_9$H). Gibbs free energy of hydrogenation curves for Co$_2$Si$_2$O$_9$H$_n$ are similar to Fe$_2$Si$_2$O$_9$H$_n$ except that the $\mu_H$ at the transition point for  Co$_2$Si$_2$O$_9$H$_n$ is slightly smaller. 

\subsubsection{Ni-silicates}

In Table \ref{table:ni-stability} we observe stability trends that are very similar to Table \ref{table:fe-stability}. Again, the stability increases with increased hydrogenation and hull energies of trioctahedral Ni-silicates are consistently smaller than dioctahedral derivatives. However, the hull energy of Ni$_2$Si$_2$O$_9$ is larger compared to Fe$_2$Si$_2$O$_9$, whereas the hull energy of Ni$_3$Si$_2$O$_9$ is smaller than Fe$_3$Si$_2$O$_9$. This indicates that overall, Ni-silicates have a stronger tendency to form trioctahedral compounds compared to Fe-silicates. This is to be expected: Ni commonly has an oxidation state of 2+ unlike Mn, Fe and Co which are more commonly found in 2+ and 3+ oxidation states. Similarly, we have found that the only stable Cr-silicate in our work is Cr$_2$Si$_2$O$_9$H$_4$, where Cr has an oxidation state of 3+ in the dioctahedral form. In Ni$_3$Si$_2$O$_9$H$_4$, Ni prefers an oxidation state of 2+  in the trioctahedral form. Hence, it can be argued that for Mn-Co silicates dioctahedral and trioctahedral phases are in competition and coexist, but for Cr and Ni silicates one phase is clearly favored over the other. 

\begin{table}[h]
\caption{ Ground state properties of 2D Ni-silicates \label{table:ni-stability}}

\begin{tabular}{l c c c c c}
\hline
\hline
Material & $N_{ox}^{Ni}$ & EC & E$_f$ & E$_{hull}$ & $\mu_M$\\
\hline
\multicolumn{6}{l}{\textit{Nontronites}} \\
Ni$_2$Si$_2$O$_8$	    & 4+    & d$^6$         & 	-1.709	&	0.256	& 1.55 \\
Ni$_2$Si$_2$O$_8$H	    & 3.5+  & d$^6$, d$^7$  & 	-1.761	&	0.187	& 1.56 \\
Ni$_2$Si$_2$O$_8$H$_2$	& 3+    & d$^7$         & 	-1.797	&	0.135	& 1.47 \\
\hline
\multicolumn{6}{l}{\textit{Kaolinites}}	\\													
Ni$_2$Si$_2$O$_9$	    & 5+    & d$^5$         & 	-1.511	&	0.304	&	1.29	\\
Ni$_2$Si$_2$O$_9$H	    & 4.5+  & d$^5$, d$^6$  &	-1.579	&	0.264	&	1.61	\\
Ni$_2$Si$_2$O$_9$H$_2$	& 4+    & d$^6$         & 	-1.571	&	0.232	&	2.12	\\
Ni$_2$Si$_2$O$_9$H$_3$	& 3.5+  & d$^6$, d$^7$  &	-1.671	&	0.170	&	2.14	\\
Ni$_2$Si$_2$O$_9$H$_4$	& 3+    & d$^7$         & 	-1.647	&	0.110	&	2.16	\\
\hline
\multicolumn{6}{l}{\textit{Lizardites}}	\\													
Ni$_3$Si$_2$O$_9$	    & 3.33+ & d$^6$, d$^7$  &	-1.756	&	0.069	&	1.33	\\
Ni$_3$Si$_2$O$_9$H	    & 3+    & d$^7$         & 	-1.788	&	0.030	&	1.40	\\
Ni$_3$Si$_2$O$_9$H$_2$	& 2.67+ & d$^7$, d$^8$  &	-1.792	&	0.021	&	1.54\\
Ni$_3$Si$_2$O$_9$H$_3$	& 2.34+ & d$^7$, d$^8$  &	-1.774	&	0.034	&	1.66\\
\bf{Ni$_3$Si$_2$O$_9$H$_4$}	& 2+    & d$^8$         & 	-1.804	&	0.000	&	1.80	\\
\hline
\end{tabular}
\end{table}

We studied the hydrogenation of Ni-silicates using Eq. \ref{eq:gibbs} and show the data in the supplementary information. Hydrogenation of Ni$_3$Si$_2$O$_9$H$_n$ follows the same trend that we discussed in Co$_3$Si$_2$O$_9$H$_n$. There is a stability range for Ni$_3$Si$_2$O$_9$H that exists, but it is much smaller compared to Mn$_3$Si$_2$O$_9$H and Fe$_3$Si$_2$O$_9$H. 

\subsubsection{Structural stability of kaolinites and lizardites}
It is known that the parent compounds of kaolinite Al$_2$Si$_2$O$_9$H$_4$ and Mg$_3$Si$_2$O$_9$H$_4$ exist naturally \cite{Bish1993, Mellini1987}, but our literature search has not yielded any work on theoretical phonon dispersion of these materials. Nevertheless, in order to show that transition metal counterparts of these 2D materials are structurally stable, we calculated the phonon dispersions of Cr$_2$Si$_2$O$_9$H$_4$ and Mn$_3$Si$_2$O$_9$H$_4$ as exemplars. (An exhaustive study of the phonon dispersions of all the metal silicates here is avoided due to high computational costs of performing so many calculations.) While Mn$_3$Si$_2$O$_9$H$_4$ is structurally stable, for Cr$_2$Si$_2$O$_9$H$_4$ we see very weak structural instabilities near $\Gamma$. We find that the size and presence of these instabilities depend strongly on numerical parameters such as the size of the supercell or the force-constant cutoff radius, indicating that they are numerical artifacts. A detailed discussion is provided in the SI \cite{Supp}.  Our conclusion is that these 2D silicates are structurally stable.

\subsection{Magnetic structure}
\setlength{\fboxsep}{0pt}%
\setlength{\fboxrule}{2pt}%
\begin{figure}[h]
\includegraphics[]{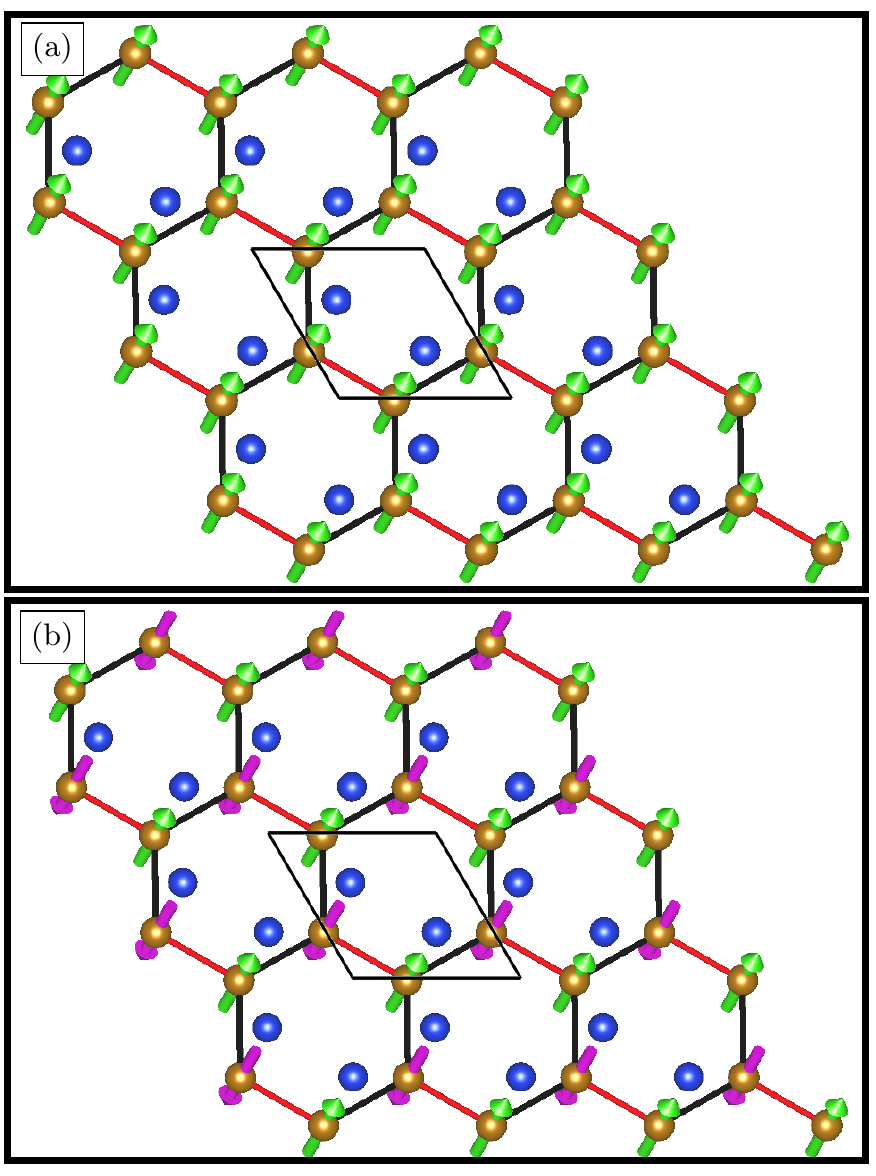}
  \caption{(a) FM and (b) G-AFM ordering considered for kaolinite structures. O and H atoms are ommited for clarity, only the transition metal (TM) atoms and Si are shown in gold and blue, respectively. Up and down spin directions are shown with green and magenta arrows, respectively. Distortions from the perfect honeycomb lattice are displayed also: longer TM bonds are shown in black, and shorter TM bonds are shown in red. The primitive cell of the structure is shown with black lines at the center of each figure.}
  \label{fgr:kao-mag}
\end{figure}

To better understand the magnetic structures, we begin with the geometry of the silicates starting with the kaolinites. In Fig. \ref{fgr:kao-mag}, we show that the kaolinites have a distorted honeycomb lattice. This distortion yields quasi-1D structures having equidistant transition metal (TM) atoms along 1D zigzag chains and a different TM-TM distance perpendicular to the zigzag chains. This is because the [SiO$_4$]$^{-2}$ layer modulates the O ligands such that they breaks the hexagonal symmetry of the transition metal oxide layer. For example, each Cr atom in Cr$_2$Si$_2$O$_9$H$_4$ has three Cr neighbors with two of them d(Cr-Cr)$_1$=3.072 {\AA} away (black lines in Fig. \ref{fgr:kao-mag}) and one d(Cr-Cr)$_2$=3.012 {\AA} away (red lines in Fig. \ref{fgr:kao-mag}). This pattern of bond lengths is important to consider when comparing to other compounds such as MnX$_3$ \cite{Sun2018} where the Mn also has a $d^4$ electronic state similar to Mn$_2$Si$_2$O$_9$H$_4$. In this work we aim to screen whether there are any AFM phases which are energetically more favorable than the FM phase. In order to understand the magnetic interactions between transition atoms with bond length alternations, such as d(Cr-Cr)$_{1,2}$, larger supercells should be considered in future work.

\setlength{\fboxsep}{0pt}%
\setlength{\fboxrule}{2pt}%
\begin{figure}[h]
\begin{subfigure}[htbp]{\linewidth}
        \begin{tikzpicture}
        \node[inner sep=0pt] 
        {\fbox{\includegraphics[width=\linewidth]{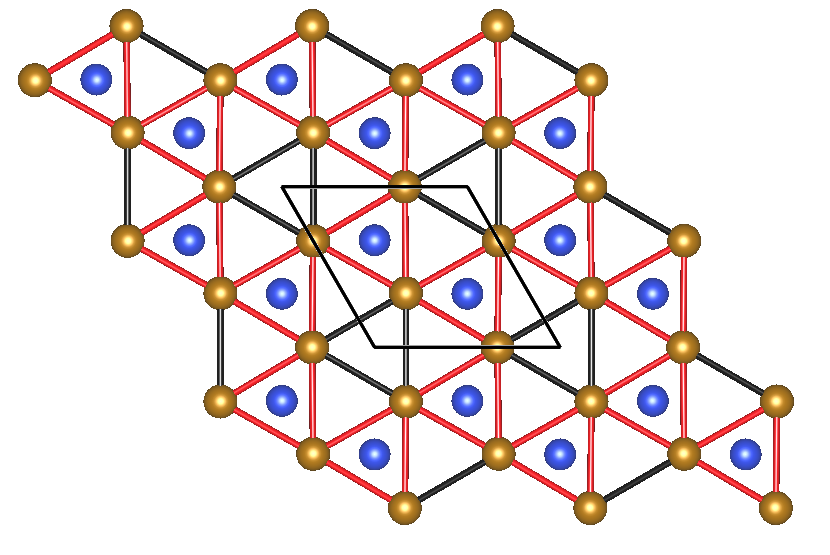}}};
        \end{tikzpicture}
\end{subfigure}
  \caption{Distortion of the triangular lattice in lizardites. Gold spheres show the TM atoms, blue spheres show Si atoms. O and H are omitted for clarity. Longer TM-TM bonds are shown using black, whereas the shorter TM-TM bonds are shown using red lines. The primitive cell of the structure is indicated using black lines in the center of the figure. }
  \label{fgr:liz}
\end{figure}

In Lizardites, a similar distortion as in the kaolinites is found. A prototype of a lizardite structure is shown in Fig. \ref{fgr:liz}. For example, each Fe atom in Fe$_3$Si$_2$O$_9$H$_4$ has six Fe neighbors, where any Fe has four Fe neighbors with d(Fe-Fe)$_1$=3.14 {\AA} (red lines in Fig. \ref{fgr:liz}) and two Fe neighbors with d(Fe-Fe)$_2$=3.19 {\AA} (black lines in Fig. \ref{fgr:liz}). This is because the Si-O layer modulates the transition metal oxide layer. Each Si-tetrahedron is centered on top of an equilateral triangle of Fe-Fe bonds with d(Fe-Fe)$_1$ as can be seen in Fig. \ref{fgr:liz}. As the Si-O layer forms a honeycomb lattice, the site above the equilateral triangles with d(Fe-Fe)$_2$ is empty, hence the Fe-Fe distances are modulated. Non-FM magnetic orderings in the primitive cell of lizardite structure are spin-frustrated. Therefore, we calculate a striped AFM ordering using a rectangular supercell which is simply doubled along one primitive vector. Although even larger supercells will allow for more complex magnetic ordering patterns, here we restrict ourselves to these simple orderings to obtain an overall assessment of the magnetic ordering energy scales at reasonble compuatational costs.  To understand the potential of these structures for 2D FM behavior, comparing simple FM and AFM orderings provides a good estimate of the relative stability.

In Table \ref{table:magnetic}, we compare antiferromagnetic (AFM) and ferromagnetic (FM) phases of kaolinites and lizardites.
\begin{table}[h]
\caption{Magnetic properties of transition metal silicates  \label{table:magnetic}}
\begin{tabular}{l c}
\hline
\hline
Material & $E_{AFM} - E_{FM}$ (meV per metal atom)\\
\hline
\multicolumn{2}{l}{\textit{Kaolinites}}	\\													
Cr$_2$Si$_2$O$_9$H$_4$	    & 3\\
Mn$_2$Si$_2$O$_9$H$_4$	    & -8\\
Fe$_2$Si$_2$O$_9$H$_4$	    & -3\\
Co$_2$Si$_2$O$_9$H$_4$	    & -28\\
Ni$_2$Si$_2$O$_9$H$_4$	    & -201\\
\hline
\multicolumn{2}{l}{\textit{Lizardites}}	\\													
Mn$_3$Si$_2$O$_9$H$_4$	    & -24\\
Fe$_3$Si$_2$O$_9$H$_4$	    & -1\\
Co$_3$Si$_2$O$_9$H$_4$	    &  1\\
Ni$_3$Si$_2$O$_9$H$_4$	    & -2\\
\hline
\end{tabular}
\end{table}

Our results in Table \ref{table:magnetic} show that the energy differences between the AFM and FM phases are mostly quite small, and that the ground state is mainly AFM.  For many, the small AFM-FM energy difference means no magnetic ordering is likely at room temperature; for Co$_2$Si$_2$O$_9$H$_4$, Mn$_3$Si$_2$O$_9$H$_4$ and Ni$_2$Si$_2$O$_9$H$_4$ the AFM order is expected to survive at or near room temperature.  We note note that  Ni$_2$Si$_2$O$_9$H$_4$ in Table \ref{table:magnetic} is a thermodynamically unstable structure since each Ni atom has a 3+ oxidation state. Similarly, Cr$_3$Si$_2$O$_9$H$_4$ is also a thermodynamically unstable material and given then Cr atoms would have a 2+ oxidation state in this structure. In the supplementary material \cite{Supp}, we show that the same conclusion can also be made for Cr$_3$Si$_2$O$_9$H$_4$. 

The main FM candidate in Table \ref{table:magnetic} is the kaolinite Cr$_2$Si$_2$O$_9$H$_4$, although the Curie temperature is likely to be very low. This is similar to the magnetism in Cr-Ni pyroxenes which yields an AFM ground state for Mn to Fe-pyroxenes \cite{Streltsov2008,Redhammer2009} but a FM ground state for Cr-pyroxenes. Pyroxenes and kaolinites are structurally rather similar. In both structures the magnetism is mediated over M-O-M bonds which are close to 90$^\circ$. In pyroxenes, however, M-O octahedra form one dimensional chains which are separated by alkali atoms such as Li and Na, as opposed to the two-dimensional M-O layer in kaolinites. In kaolinites however, the structural modulation seen in Fig. \ref{fgr:kao-mag} also indicates that the magnetic coupling is not isotropic, hence the situation is similar to the pyroxenes.
In Cr-pyroxenes, it was shown that the AFM $t_{2g}$-$t_{2g}$ exchange interaction is nearly compensated by the FM $t_{2g}$-$e_g$ exchange, but fine-tuning of these interactions is possible via the size of the alkali atom \cite{Streltsov2008}.
In NaCrGe$_2$O$_6$, the largest Cr-Cr separation was observed which yields a FM structure \cite{Streltsov2008}. Although not examined here, future work can examine if it is possible to incorporate additional transition metal atoms in the vacancies of the honeycomb lattice of Cr-kaolinite to increase FM coupling and/or have a ferrimagnetic ground state with a net magnetization.

As we previously discussed, greenalite, Fe$_{2,3}$Si$_2$O$_9$H$_4$, is observed to be ferromagnetic in the plane with a intraplane magnetic couling constant of 15 K \cite{Coey1982}. Fe sites in greenalite are disordered however, meaning that some octahedral sites are filled with Fe, whereas others are hollow. In this perspective, greenalite can be considered a solid solution of the ordered Fe-kaolinite and Fe-lizardite phases we studied. However, our calculations do not yield a ferromagnetic order in these  ordered lizardite and kaolinite phases. For Fe-silicates, our calculations yield a G-type checkerboard AFM phase for Fe$_{2}$Si$_2$O$_9$H$_4$ and striped AFM phase for Fe$_{3}$Si$_2$O$_9$H$_4$ as energetically more favorable compared to the FM phases. Therefore, we can understand that the ferromagnetism of greenalite is correlated to the disorder in its structure. Similar accounts of enhanced ferromagnetism with increased disorder were reported in 3D alloys \cite{Berciu2001} as well as disordered and doped 2D-materials \cite{Shi2013, Mishra2013}. As we predict both lizardite and kaolinite Fe-silicates to be thermodynamically stable, synthesis conditions might be engineered to make partially occupied Fe-O layer which can yield magnetic properties similar to greenalite.

\subsection{Piezoelectric properties}
Here, we describe the piezoelectric properties of the thermodynamically stable 2D silicates identified above. In order for a material to be piezoelectric, it has to be insulating or semiconducting plus have broken inversion symmetry. We provide the density of states plots and the band gaps of the kaolinites and lizardites we studied in the SI \cite{Supp}: band gaps of the transition metal silicates we studied are all above 1 eV \cite{Supp}. Both kaolinites and the lizardites have symmetry point groups that do not include inversion, hence they are expected to have a finite piezoelectric response under strain or electric field. As the side views of Fig. \ref{fgr:top_view}(d)-(f) show, the Si-O and transition metal-O layers are chemically stacked and the resulting dipole is in the $z$-direction.  We use the following standard relations to calculate the elastic modulus tensor $C_{ij}$ and piezoelectric strain tensor $e_{\alpha j}$:
\begin{equation}
\begin{gathered}
    C_{ij} = \frac{d\sigma_{i}}{d\eta_{j}}\,, \\
    e_{\alpha{j}} = \frac{dP_{\alpha}}{d\eta_{j}} = e_{\alpha{j}, c} + e_{\alpha{j}, i} \\
    e_{\alpha{j}, c} = \left.\frac{\partial P_{\alpha}}{\partial\eta_{j}}\right\rvert_{u} = \left.\frac{\partial^2E}{\partial\mathcal{E}_{\alpha}\partial\eta_{j}}\right\rvert_{u}\,, \\
    e_{\alpha{j}, i} = \sum_m\frac{{\partial}P_{\alpha}}{{\partial}u_{m}}\frac{{\partial}u_{m}}{{\partial}\eta_{j}} \,,\\
    d_{\alpha{j}} = \frac{dP_{\alpha}}{d\sigma_{j}} = \sum_i e_{\alpha{i}}\cdot  (C^{-1})_{ji}\,.
\end{gathered}
\label{eq:piezo}
\end{equation}
Here, $\sigma_i$ is the stress tensor, $\eta_{j}$ is the strain tensor, $e_{\alpha{j}}$ is the piezoelectic tensor with subscripts $i$ and $c$ denoting the ionic and clamped ion components respectively, $E$ is the total energy of the system,  $\mathcal{E}_{\alpha}$ is the imposed electric field vector, $u$ are the atomic displacements from equilibrium, $d_{\alpha{j}}$ is the piezoelectric strain tensor, and $P_{\alpha}$ is the polarization vector.  Greek indices such as $\alpha$ represent axis directions and Latin indices such as $i,j$ describe tensor components using Voigt notation.

\begin{table*}[ht]
\caption{Elastic coefficients ($C_{ij}$) and piezoelectric coefficients ($e_{ij}$ and $d_{ij}$) of 2D kaolinites.  By symmetry, $C_{12} = C_{21}$ and $e_{11}=e_{12}=0$. \label{table:piezo_kao}}
\begin{tabular}{l c  c c c c c c c c c c c c c c c c c c c}
\hline
& \multicolumn{4}{c}{\textit{clamped-ion}} &&& \multicolumn{14}{c}{\textit{relaxed-ion}}\\
\cline{2-5} \cline{8-21}
&  $e_{21}$ & $e_{22}$ & $e_{31}$ & $e_{32}$  &&& $C_{11}$ & $C_{12}$ & $C_{22}$ & $C_{66}$ & & $e_{21}$ & $e_{22}$ & $e_{31}$ & $e_{32}$ & & 
$d_{21}$ & $d_{22}$ & $d_{31}$ & $d_{32}$ \\
& \multicolumn{4}{c}{pC/m} &&& \multicolumn{4}{c}{N/m} && \multicolumn{4}{c}{pC/m} && \multicolumn{4}{c}{pm/V}\\
 \cline{2-5} \cline{8-11} \cline{13-16} \cline{18-21} 
Cr$_2$Si$_2$O$_9$H$_4$     & -6.5	& 	-3.5	& 	26.0	& 	38.1 & & & 
148.9 & 54.4 & 132.2 & 32.8 & &
40.9	& 	68.9	& 	-39.3	&	-23.9 & & 
0.10 & 0.48 & -0.23 & -0.08\\
Mn$_2$Si$_2$O$_9$H$_4$     &  1.6	& 	-6.7	& 	30.7	& 	41.2  & & & 
141.1 & 49.8 & 125.7 & 25.8 & & 
83.8	& 	82.3	& 	-54.8	& 	-36.9  & & 
0.42 & 0.49 & -0.33 & -0.16\\
Fe$_2$Si$_2$O$_9$H$_4$     &  -7.6	& 	-1.6	& 	24.4	& 	40.3 & & & 
143.4 & 52.9 & 123.9 & 29.3 & & 
41.4	& 	132.0	& 	-40.2	& 	-21.0 & &
0.12 & 1.11 & -0.26 & -0.06\\
Co$_2$Si$_2$O$_9$H$_4$     &  6.9	& 	-4.7	& 	27.2	& 	38.1 & & & 
141.5 & 48.9  & 117.8 & 23.8  & &
53.7	& 	53.8	& 	-40.9	& 	-20.2 & & 
0.26 & 0.35 & -0.27 & -0.06\\
Ni$_2$Si$_2$O$_9$H$_4$     & 1.2	& 	8.2	& 	28.4	& 	39.8 & & &
138.2 & 48.3 & 113.8 & 29.3 & &
83.1	& 	56.3	& 	-45.2	& 	-24.1 & & 
0.50 & 0.29 & -0.29  & -0.10\\
\hline
\end{tabular}
\end{table*}

\begin{table*}[ht]
\caption{Elastic coefficients ($C_{ij}$) and piezoelectric coefficients ($e_{ij}$ and $d_{ij}$) of 2D lizardites.  By symmetry ${C}_{12} = C_{21}$,  $C_{11} = C_{22}$, $e_{21} = e_{22}$, $e_{31} = e_{32}$, and $e_{11} = e_{12}=0$.
\label{table:piezo_liz}}
\begin{tabular}{l c c c c c c c c c c c c }
\hline
& \multicolumn{2}{c}{\textit{clamped-ion}}  & & \multicolumn{9}{c}{\textit{relaxed-ion}}\\
\cline{2-3}\cline{5-13}
          & $e_{22}$ & $e_{32}$  & & $C_{12}$ &  $C_{22}$ & $C_{66}$ & & $e_{22}$ & $e_{32}$ & & $d_{22}$ & $d_{32}$\\

 & \multicolumn{2}{c}{pC/m}         & & \multicolumn{3}{c}{N/m}        & & \multicolumn{2}{c}{pC/m} & & \multicolumn{2}{c}{pm/V} \\
\cline{2-3} \cline{5-7} \cline{9-10} \cline{12-13} 
Mn$_3$Si$_2$O$_9$H$_4$     & 0.6    & 40.2      & & %
139.7 & 54.9  & 42.0  & & %
69.4 & -21.5 & & %
0.82 & -0.11 \\
Fe$_3$Si$_2$O$_9$H$_4$     &  2.1	& 	42.9	& & %
139.0 & 55.3  & 41.7 & &  %
12.8 & -35.7 & & %
0.15 & -0.18 \\
Co$_3$Si$_2$O$_9$H$_4$     &  3.1	& 	41.5    & & %
148.0 & 62.3  & 43.8 &  &  %
39.3 & -27.2 & & %
0.46 & -0.13 \\
Ni$_3$Si$_2$O$_9$H$_4$     &  4.9	& 	36.3    & &        %
161.8 & 64.7  & 48.1 &  & %
32.2 & -35.8 & & %
0.33 & -0.16 \\
\hline
\end{tabular}
\end{table*}

We apply a symmetry and dimensionality analysis to define in-plane directions and independent components of the elastic and piezoelectric tensors. Plane-wave based DFT codes such as VASP calculate the $C_{ij}$ and $e_{\alpha j}$ constants based on periodic boundary conditions of a 3D system. Therefore, it is important that these quantities are modified or converted for a 2D case with in-plane stress and strain. For a 2D system, this means that the $\sigma_i$ and $\epsilon_{j}$ are zero when $i$ or $j$ involves the out-of-plane $z$ direction \cite{Blonsky2020, Duerloo2012}. Also, a renormalization is needed for the elastic and strain tensors such that $C_{ij}^{2D} = a_z\cdot C_{ij}^{3D}$ and $e_{ij}^{2D} = a_z\cdot e_{ij}^{3D}$, where $a_z$ is the length of the simulation cell in $z$-direction.
However, the polarization $P_\alpha$ is not restricted to remain in-plane. 

We use orthorhombic simulation cells for both lizardite and kaolinite derivatives to calculate elastic and piezoelectric constants as defined in the SI \cite{Supp}. In these cells, the in-plane lattice parameters along the $x$-axes were chosen to be longer than the lattice parameters on $y$-axes. 
In all structures, the $z$-direction is perpendicular to the $xy$-plane. Lizardites have the $3m$ point-group symmetry, hence $x$ and $y$ in-plane directions are identical \cite{Duerloo2012, Blonsky2020}. However, the honeycomb lattice of kaolinites leads to an anisotropy between $x$ and $y$ directions, which was previously noted by Sato et. al. \cite{Sato2005}.
Since 2D kaolinite and lizardite crystals have the $m$ and $3m$ point group symmetries respectively, the complete piezoelectric strain tensor can be obtained using only the independent tensor elements. For the $3m$ point group symmetry, these are $e_{22}=e_{21}=e_{16}$ and $e_{32}=e_{31}$ and in the $m$ point group symmetry these are $e_{21}$, $e_{22}$, $e_{31}$, $e_{32}$ and $e_{16}$ \cite{Gallego2019}. In both structures, $e_{11} = e_{12} = 0$. 

In Tables \ref{table:piezo_liz} and \ref{table:piezo_kao}, we show that the elastic properties of transition metal silicates are very similar to each other. The main difference is that elastic constants of kaolinites are smaller than the lizardites. This is most likely because the hexagonal vacancy in kaolinites  allows additional room for relaxation, leading to smaller elastic moduli. Indeed, a similar conclusion can be made using the bulk forms of kaolinite and lizardite (3D Al$_2$Si$_2$O$_9$H$_3$ and Mg$_3$Si$_2$O$_9$H$_3$, respectively) where $C_{11}$ elastic constants of 200 and 245 GPa were calculated using DFT, respectively \cite{Sato2005, Reynard2007}. Additionally, the $C_{11}$ elastic constant of kaolinites tend to decrease from Cr- to Ni-kaolinite, whereas in lizardites the trend is the opposite. This can be related to the trends in structural parameters. We find that the average volume of the transition metal octahedra and the in-plane surface area decreases going from Mn to Ni in both structures. This correlates with the reduced atomic size going towards Ni, hence more tightly packed structures and larger elastic constants for the lizardites. 

In Tables \ref{table:piezo_liz} and \ref{table:piezo_kao}, piezoelectric and elastic tensor components are reported.  The general trends found in these tables are: In both structures, the  clamped-ion and relaxed ion piezoelectric constants, $e_{ij}$, differ dramatically and lead to a change of sign in $e_{31}$ and/or $e_{32}$. 
We should mention that the sign of the $e_{31}$ and $e_{32}$ piezoelectric constants depend on the orientation of the 2D layer, i.e. whether the silicate layer is on top of the transition metal oxide layer. We kept the orientation of layers fixed in all our calculations such that transition metal oxide layer is always on top. 
This change of sign has been also observed in the $e_{33}$ constants of 3D auxetic piezoelectric crystals and vdW solids \cite{Liu2017a, Liu2020},  and it was found that the magnitude of the ionic contribution is typically much larger than the magnitude of the clamped-ion contribution in these auxetic materials. In ref. \citenum{Liu2020}, it was pointed out that this large ionic contribution is a main character of auxetic piezoelectric materials, and both $e_{31}$ and $e_{33}$ constants are negative. In quasi-2D materials, $e_{33}$ constants can be obtained experimentally \cite{Cui2018}, but this is challenging for the computational methods. 
A standard ab initio calculation of an isolated quasi-2D layer in vacuum must yield $e_{33}=0$ since stretching the simulation cell along the z direction is equivalent to adding vacuum to the simulation which doesn’t modify the 2D material in any way.  
To try to extract a value of $e_{33}$ that corresponds approximately to a value that might be obtained by an indentation experiment, we compute $e_{33}$ for the bulk structure (stacked 2D layers along the $z$-direction). This simplified approach uses the 2D layer as both substrate and indenter, which should give a good order of magnitude estimate and also relative ordering between the $e_{33}$ constants of 2D layers with different cation substitutions. Consult the SI \cite{Supp} for additional details.

Similar to $e_{31}$ and $e_{32}$ constants, we find that the  $e_{22}$ constants in both materials differ significantly between the clamped-ion and relaxed-ion conditions. However, similar observation for the in-plane piezoelectric constants were previously made on a diverse set of 2D materials and found to be related to their mechanical softness and ionic relaxation in the out-of-plane direction \cite{Sevik2016}. In kaolinites, we find that $d_{31}$ are larger than $d_{32}$ constants in magnitude, which can be correlated to a similar difference in $e_{31}$ and $e_{32}$ constants as a result of the anisotropy observed in these materials. Similarly, $d_{32}$ constants in kaolinites are typically smaller than that of lizardites, however, $d_{31}$ constants in each material (in lizardites $d_{31}=d_{32}$) are comparable. The $d_{3j}$ constants in tables \ref{table:piezo_liz} and \ref{table:piezo_kao} are all smaller than 1 pm/V, which is on par with most quasi-2D materials which all have $d_{31}$ smaller than 1 pm/V, such as Janus-type transition metal dichalcogenides \cite{Hinchet2018}, buckled hexagonal III-V compounds \cite{Blonsky2020}, and doped graphene \cite{Ong2012}.

We find that the main contribution to the relaxed ion out-of-plane piezoelectric constants, $e_{3j, i}$, comes from the displacements of the Si and O atoms in the SiO$_4$ tetrahedra in these structures, which explains relatively similar performance across different substituents \cite{Supp}. We analyze the atomic displacements due to strain that contribute to $e_{\alpha j, i}$ using the tensor $A_{mj} =
\partial u_m / \partial \eta_j$ in the SI \cite{Supp}.
We find that the SiO$_4$ tetrahedra move closer to the MO$_6$ layer with tensile strain as would be expected from a material with a positive Poisson ratio. Strain in the $x$ ($\equiv1$) direction induces displacements in the Si-O bond between the SiO$_4$ tetrahedra and the MO$_6$ layer such that Si and O atoms in the SiO$_4$ tetrahedra move up or down collectively. The magnitude of displacements in these atoms are larger compared to the rest of the system. 

\section{Conclusion}

In conclusion, we have presented a detailed theoretical investigation of the thermodynamic, electronic, magnetic and piezoelectric properties of 2D transition metal silicates, M$_{2-3}$Si$_2$O$_9$H$_n$, where $n$=(0,4). We show that these materials can be thermodynamically stable with hydrogenation. The symmetry of these structures dictates that a finite piezoelectric response exists, and we find that it is on par with a wide range of quasi-2D materials that show similar performance. Our long-term goal is to create a material that can possess ferromagnetism (ideally at elevated temperatures) and also has piezoelectric properties, so that the two can be coupled to each other to modulate the magnetic properties. Future studies are needed to compute the coupling of magnetic and piezoelectric properties to the external stimuli such as electric field or strain. Although we find that the magnetic properties of these materials are predominantly antiferromagnetic or weakly ferromagnetic at best, we expect that this materials framework and the facile experimental synthesis methods \cite{Zhou2019} will allow further engineering of the transition metal layer with a richer chemical phase space and improved possibilities for magnetic ordering.  

\section*{Acknowledgement}
We acknowledge the Army Research Office grant W911NF-19-1-0371 for the funding of this work and also the computational resources provided by the institutional clusters at Yale University.
\bibliography{main}

\end{document}